\documentstyle[twocolumn]{article}
\begin{document}
\setlength {\textwidth} {18.2cm} \setlength {\oddsidemargin}
{-0.5cm}\setlength  {\evensidemargin} {-0.5cm} \setlength
{\topmargin} {-3.0cm}
\setlength  {\textheight} {26.5cm}

\setlength{\textwidth}{18cm}
\newcommand{\ie}{{\em i.e. }}
\newcommand{\al}{{\em et al }}
\newcommand{\eg}{{\em e.g. }}
\newcommand{\FIXME}{{\bf FIXME}}
\newcommand{\pfc}{{$\mbox{PF}\overline{\mbox{C}}$}}
\newcommand{\pfs}{{$\mbox{PF}\overline{\mbox{S}}$}}

\title{FSSP to SCOP and CATH (F2CS) Prediction Server\vspace{4cm}}
\author{Gad Getz$^1$, Alina Starovolsky$^2$ and Eytan Domany$^1$ \\
$^1$Department of Physics of Complex Systems, Weizmann Institute
of Science, Rehovot 76100, Israel\\
$^2$Computer Science Department, Ben-Gurion University of the
Negev, Beer-Sheva 84105, Israel}
\date{\today}
\maketitle


\noindent
{\bf ABSTRACT} \\
{\bf Summary:} The F2CS server provides access to the software,
F2CS2.00, that implements an automated prediction method of SCOP
and CATH classifications of proteins, based on their FSSP Z-scores
(Getz {\it et al.},
2002),\\
{\bf Availability:} Free, at \\
{\small http://www.weizmann.ac.il/physics/complex/compphys/f2cs/}. \\
{\bf Contact:} eytan.domany$@$weizmann.ac.il \\
{\bf Supplementary information:} The site contains links to
additional figures and tables.
\vspace{0.3cm}

Since during evolution protein structures are much more conserved
than sequences and even functions \cite{holm96}, proteins are
usually classified first by their structural similarity. Newly
solved structures of proteins are regularly stored in the Protein
Data Bank (PDB) \cite{bernstein77}. Many research groups study the
diversity of protein structures and maintain web-accessible
hierarchical classifications of them. Three widely used databases
are FSSP \cite{holm97}, CATH \cite{orengo97} and SCOP
\cite{loconte00}; although each has its own way to compare and
classify proteins, the resulting classification schemes are,
largely, consistent with each other \cite{getz02,getz98,hadley99}.

The major difference between these three classification schemes,
relevant to this work, is their degree of automation. FSSP is
based on a fully automated structure comparison algorithm, DALI
\cite{holm94,dietmann01}, that calculates a structural similarity
measure (represented in terms of Z-scores) between pairs of
structures of protein chains taken from the PDB. FSSP first
selects a subset of representative structures from the PDB and
then applies the DALI algorithm to calculate the Z scores for all
pairs of representatives. Next, they calculate the Z scores
between each representative and the PDB structures it represents.
Being fully automated, FSSP can be updated fairly often. FSSP was
recently extended by a new database, called Dali \cite{holm03},
which contains all-against-all Z-scores between chains and domains
of a larger representative set, PDB90 \cite{hubbard99}, in which
no two chains are more than 90\% sequence identical. In contrast,
CATH and SCOP use manual classification at certain levels of their
hierarchy, which slows down the classification process and makes
it more subjective and error-prone.

CATH arranges protein domains in a four-level hierarchy according
to their {\bf C}lass (secondary structure composition), {\bf
A}rchitecture (shape formed by the secondary structures), {\bf
T}opology (connectivity order of the secondary structures) and
{\bf H}omologous superfamily (structural and functional
similarity). Classification of Architecture is done by visual
inspection; hence CATH is partially manual.

The top level ({\bf C}lass) of the SCOP database also describes
the secondary structure content of a protein domain. The next
level ({\bf F}old) groups together structurally similar domains.
The lower two levels (superfamily and family) describe near and
distant evolutionary relationships \cite{levitt76}. "Fold" largely
corresponds to CATH's topology level \cite{getz02}. SCOP is
constructed manually, based on visual examination and comparison
of structures, sequences and functions.

We present here a web-based server, available at {\small
http://www.weizmann.ac.il/physics/complex/compphys/f2cs/}, whose
aim is to predict, without human intervention, using a protein's
FSSP (or DALI) Z-scores, it's full SCOP and CATH classifications.
This can help classify proteins of known structure that were not
yet processed by SCOP or CATH (whose new releases are provided
about every 6 months), and call attention to yet unseen structural
classes.

If a protein appears in FSSP, the server returns our prediction.
If it is not in FSSP, the user can submit the new structure to the
DALI server, insert the resulting Z-scores into our server and
obtain its predicted classification. In both cases F2CS outputs a
table showing the prediction, along with its confidence level.

\begin{table*}[tbh]
{\small{
\begin{tabular}{ |l|l|l|l|l|l|l|l|l|l|l|l|l|} \hline

Chain id & \multicolumn{4}{|c|}{CATH v2.5}
&\multicolumn{3}{|c|}{{\bf CATH Prediction}} &
\multicolumn{3}{|c|}{SCOP 1.63} & \multicolumn{2}{|c|}{{\bf SCOP
Prediction}} \\

\cline{2-13}
 & \# & C & A & T & {\bf C} &{\bf A} & {\bf T} & \# & C & F & {\bf C} & {\bf F} \\

\hline

1dowb & -1 &  &  &  & {\bf 1} & {\bf 20} & {\bf 5} & -1 &  &
& 8 & 1 \\

\hline

$\underline{\mbox{Success\%}}$ & & & & & 100 & 99 & 100 & & & & 97
& 100
\\\hline
\end{tabular}
}} \vspace{0.2cm}

{\small{\noindent {\bf Table 1:} Results obtained by submitting
``1dowb'' to the F2CS server. This protein was classified by
neither CATH v2.5 nor SCOP 1.63 (indicated by -1 in the "number of
domains" columns). We predict the following classifications:
1.20.5 for CATH and 8.1 for SCOP, both at 100\% confidence level.}
\label{tab:1dowb} }
\end{table*}

\vspace{0.3cm}
\noindent {\bf THE SERVER}

\noindent The current predictions are based on the latest versions
of the databases; FSSP (Jun 16, 2002 update), combined with the
Dali database (preliminary version, May 2003); CATH version 2.5
(Jul 2003) and SCOP 1.63 release (May 2003). The FSSP database
contains 27182 chains, 2860 out of which are representatives. We
superimposed on these the Z-scores from the Dali database, which
were calculated for 6433 PDB90 chains; we refer to the combined
database as FSSP/DD. Only significant Z-scores are reported
($\ge2$) and used; all other Z-scores are assumed to be zero.

The server implements our method \cite{getz02}, {\em
Classification by Optimization} (CO), an optimization procedure
that searches for that class assignment of proteins (that were not
yet processed by CATH or SCOP), which attains a minimal cost. The
cost of an assignment is the sum of Z-scores between all pairs of
proteins that were not assigned to the same class. This is a
"partially supervised"  algorithm, since it utilizes for its
prediction the labels of the proteins with known classification
and also the Z scores among the training and predicted sets. We
can not classify "isolated" proteins, which are not connected by a
path of neighboring chains ({\it i.e.} $Z\ge2$) to a chain of
known classification.

We generate a prediction database of chains which appear in
FSSP/DD but not in SCOP or CATH by applying our algorithm for each
classification scheme.
The FSSP/DD version we are using contains 4014 chains which do not
appear in CATH v2.5 (we supply a prediction for 3170 of these) and
511 which are not in SCOP 1.63 (for 403 of these we have a
prediction); 272 chains appear in neither CATH nor SCOP. Since
CATH and SCOP handle protein domains whereas FSSP/DD entries are
protein chains (consisting of one or more domains\footnote{We do
not classify the few cases, when a single domain contains several
different chains or a combination of their parts.}), we use as a
training set the single domain chains that are of known
classification. Note that SCOP and CATH do not always agree on
their separation of proteins into domains.
%
%


Our prediction's success rate was estimated using a blind test in
which we hid the assignments of 3605 proteins from CATH and 4570
proteins from SCOP and tested our predictions against the known
classifications. The success rate was tested for each class
separately (see website for details). Due to larger number of
training examples and more stringent criteria for attempted
classification, the success rate has improved over our previous
work.


With every new release of the databases, F2CS can be updated; the
newly released CATH/SCOP classifications are added to the training
set, while predictions are made for proteins contained in a new
FSSP/DD release which are not yet classified by CATH or SCOP.

\vspace{0.3cm}

\noindent {\bf USAGE}

\noindent In order to retrieve our prediction for CATH's class,
architecture and topology or SCOP's class and fold of a protein,
enter the protein chain's identifier in the search box and submit
the query. If the protein appears in our database, a table will be
returned containing both the known and the predicted SCOP and CATH
classifications. For example, submission of the chain identifier
``1dowb'', which was classified neither by CATH v2.5 nor SCOP
v1.63, returns Table~1. We predict CATH classification 1.20.5 and
SCOP 8.1, both near 100\% confidence level. The "Success\%" link
points to a table with the exact numbers by which the success
rates were estimated.

In case the queried protein is not in our database, the user can
obtain its predicted classification by following these two steps:
(a) submit the protein's PDB file to the DALI server (the engine
behind FSSP) which calculates its structural similarity to the
FSSP representatives and returns a list of the representatives and
Z-scores for which $Z\ge2$. (b) Paste DALI's reply in the
appropriate query box in our server.


\vspace{0.3cm}

\noindent   {\bf ACKNOWLEDGEMENTS}

We thank L.~Holm for directing us to her new Dali database, and
M.~Vendruscolo for his advice and active involvement in the
initial stages of this project, which was partially supported by
the German-Israel Science Foundation (GIF). G.G. is supported by
the Sir Charles Clore Doctoral Scholarship.

\vspace{0.3cm}

\noindent   {\bf REFERENCES}
{\small

}

\end{document}